\documentclass[onecolumn, 12pt, times]{elsarticle}
\usepackage{subfig}
\usepackage[utf8]{inputenc}
\usepackage[version=4]{mhchem}
\usepackage[export]{adjustbox}
\usepackage[T5,T1]{fontenc}
\usepackage{tgheros}
\usepackage{bm}
\usepackage{siunitx}
\usepackage{booktabs}
\usepackage{caption}
\usepackage{subcaption}
\usepackage{tikz}
\usetikzlibrary{shapes,arrows}

\begin{document}
\begin{frontmatter}
\title{LES-FGM modelling of non-premixed auto-igniting turbulent hydrogen flames including preferential diffusion}
\date{\today}

\author{A. Ballatore\corref{cor1}}

\author{D. Quan Reyes}
\author{H. Bao}

\author{J.A. van Oijen}

\cortext[cor1]{Corresponding author. Tel: +39 3485228436. E-mail address: a.ballatore@tue.nl (A. Ballatore)}

\affiliation{organization={Eindhoven University of Technology, Department of Mechanical Engineering, Power \& Flow}, 
                 addressline={Building 15, Groene Loper 5},
                 postcode={5612 AE}, 
                 city={Eindhoven}, 
                 country={The Netherlands}}

\begin{abstract} 

Tabulated chemistry methods are a well-known strategy to efficiently store the flows thermochemical properties. In particular, the Flamelet-Generated Manifold (FGM) is a widely used technique that generates the database with a small number of control variables. In order to build such a manifold, these coordinates must be monotonic in space and time. However, the high diffusivity of hydrogen can prevent such requisite. There have been many studies that successfully included non-unity Lewis effects in FGM, but mostly in the context of premixed flames. The problem of accounting for differential diffusion in purely non-premixed auto-igniting hydrogen flames still has to be investigated thoroughly. To avoid the non-monotonicity of control variables (the progress variable, in particular), one practical workaround is to perform the tabulation on zero-dimensional (0D) reactors rather than on one-dimensional (1D) flamelets. Various works already implemented and tested such 0D-based manifold, but mainly in the context of spray engines, where most of the composition is lean and information past the flammability limit is not relevant. The present work aims at investigating, for the first time, the applicability of a tabulation based on homogeneous reactors to study auto-igniting turbulent hydrogen jets. 

Three different techniques to extrapolate the data beyond the flammability limit are evaluated in 1D simulations and assessed against detailed chemistry results. It is shown that a combined use of homogeneous reactors at the lean side and an extrapolation with 1D flamelets on the richer side is required to capture both chemistry and diffusive effects accurately in pure hydrogen flames. Then, this manifold is coupled to Large-Eddy Simulation (LES) of three-dimensional turbulent mixing layers and evaluated against direct numerical simulation with detailed chemistry. Good agreement is found, in terms of both ignition delay and the following steady-state burning process. Further analyses are carried out on statistics and modelling. In particular, the sensitivity of the LES solution to filter width, turbulence-chemistry interaction and multidimensional flame effects is investigated to provide new relevant insights on modelling non-premixed auto-igniting turbulent hydrogen flames.

\smallskip


\end{abstract}

\end{frontmatter}

\clearpage
%



\clearpage

\section{Introduction\label{sec:introduction}} \addvspace{10pt}

The importance of hydrogen in achieving carbon neutrality is widely known, and the development of hydrogen combustion technology is progressing further and further \cite{hanely}. Computational Fluid Dynamics (CFD) is, in such perspective, a fundamental tool to optimise development time and save costs. In order to accurately estimate the combustion of such a fuel, it is necessary to use a detailed chemical reaction mechanism. However, the associated computational cost is often too demanding. It is therefore crucial to develop reduced-order models able to correctly describe the combustion process with a high computational efficiency. 

In particular, among the tabulated chemistry methods, the 
Flamelet Progress Variable (FPV) \cite{ihme} and the Flamelet Generated Manifold (FGM) \cite{vanoijen, vanoijen2} methods are well-known strategies to efficiently store the flow thermochemical properties. Instead of solving many species and an energy equation, transport equations are solved for a few reduced variables only, using them as coordinates to retrieve all the other relevant properties which are tabulated based on one-dimensional (1D) flamelet solutions. The 1D flamelet solutions are to be parametrised as function of a handful of control variables. Typically, for non-premixed flames, the choice comes down to the mixture fraction $Z$ and the progress variable $\mathcal{Y}$. The key to a correct FGM generation is ensuring that the control variables are monotonic. For each flamelet, all the dependent thermochemical variables are stored on a structured mesh, using a multi-dimensional linear interpolation for retrieval. When solving the transport equations for the control variables, the values computed at a certain time-step are used to look-up all the other properties in the table.

The high diffusivity of hydrogen, whose Lewis number is about 0.3, prevents the progress variable, typically a linear combination of species mass fractions, to be monotonic. This problem is often avoided, especially in turbulent combustion, where it is a common practice to assume unity Lewis number \cite{vreman}, as the molecular differential diffusion is assumed negligible with respect to the transport phenomena. However, in hydrogen-enriched fuels, this leads to inaccurate results \cite{ebrahim}.

There have been many successful efforts over the last decades to include more realistic preferential diffusion effects in FGM (recent examples can be found in \cite{nithin,kai}). Nevertheless, these studies focused on premixed hydrogen flames. There are few works that used FGM to address non-premixed hydrogen/air systems (e.g. \cite{jiang,ranga}) but did not model any igniting phenomena. Hence, the problem of differential diffusion in non-premixed auto-igniting hydrogen flames with tabulated chemistry has still to be investigated thoroughly. In particular, we want to extend the work of Abtahizadeh et al. \cite{ebrahim}, that encompasses numerical investigations of preferential diffusion effects in Large Eddy Simulation of \ce{CH4}-\ce{H2} auto-igniting lifted flames, to pure hydrogen jets.

\begin{figure}[b!]
    \centering
    \includegraphics[width=0.8\linewidth]{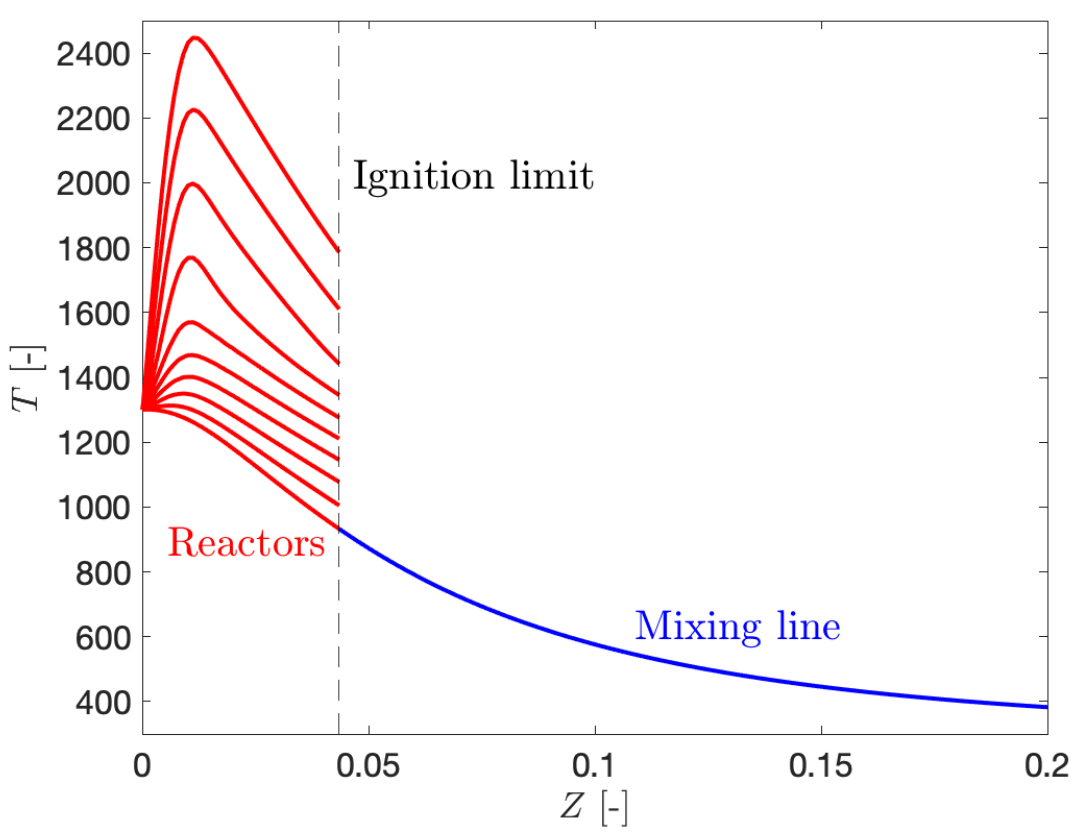}
    \caption{Temperature $T$ as function of the mixture fraction $Z$ in homogeneous reactors. After a certain $Z$ (flammability limit), the reactors do not ignite.}
    \label{reac_empty}
\end{figure}

As mentioned, the common generation of the manifold can be troublesome, as the low Lewis number of hydrogen (and \ce{H} radicals) complicates the definition of a monotonic progress variable, at least in some parts of the mixture fraction space. A practical workaround is to build the manifold using homogeneous reactors rather than one-dimensional flamelets. The table then comprises properties not affected by diffusion-advection, allowing an easier choice of the progress variable. These missing convective and diffusive effects are then accounted for when transporting $Z$ and $\mathcal{Y}$ in the CFD code. Tabulation methods based on such zero-dimensional (0D) models have already been explored in the past \cite{bekdemir,lucchini,chevillard,eguz} but not for hydrogen. These reactors do not ignite beyond a certain equivalence ratio, due to both low temperature and rich composition of the unburnt mixture, so a great part of the $Z$-range does not include any evolution from the starting mixing line (Fig.~\ref{reac_empty}). Consequently, the problem of extrapolating the data beyond such equivalence ratio must be solved, and is not trivial. A similar study is presented in literature by Ketelheun et al. \cite{anja}. However, their study is based on premixed flamelets and unity Lewis numbers, far from the real behaviour of pure hydrogen. The novelty of the present work is to conduct a similar study in non-premixed auto-igniting hydrogen flames, including preferential diffusion. For the first time, the paper at hand illustrates the ability of a tabulated chemistry based on homogeneous reactors and one-dimensional flamelets to solve the problem of non-monotonicity of the progress variable in the standard flamelet approach.

First, the procedures to construct the tables are discussed. Then, the manifolds are assessed in 1D flames against corresponding detailed chemistry results. It is shown that an extrapolation past the reactors' flammability limit is required to retain both the chemistry and the diffusivity of the richer compositions. Such method is finally coupled to Large-Eddy Simulation (LES) to investigate a three-dimensional (3D) reacting turbulent mixing layer of hydrogen and validate against Direct Numerical Simulation (DNS) computations. Further analysis are conducted to provide new relevant insights on statistics and modelling of turbulent auto-igniting hydrogen combustion. This work will serve as a basis to incorporating combustion modelling by means of tabulated chemistry in high-pressure hydrogen injections for internal combustion engines (e.g. \cite{ballatore}).

\section{Table generation}
\begin{figure}[h!]
    \centering
    \includegraphics[width=1\linewidth]{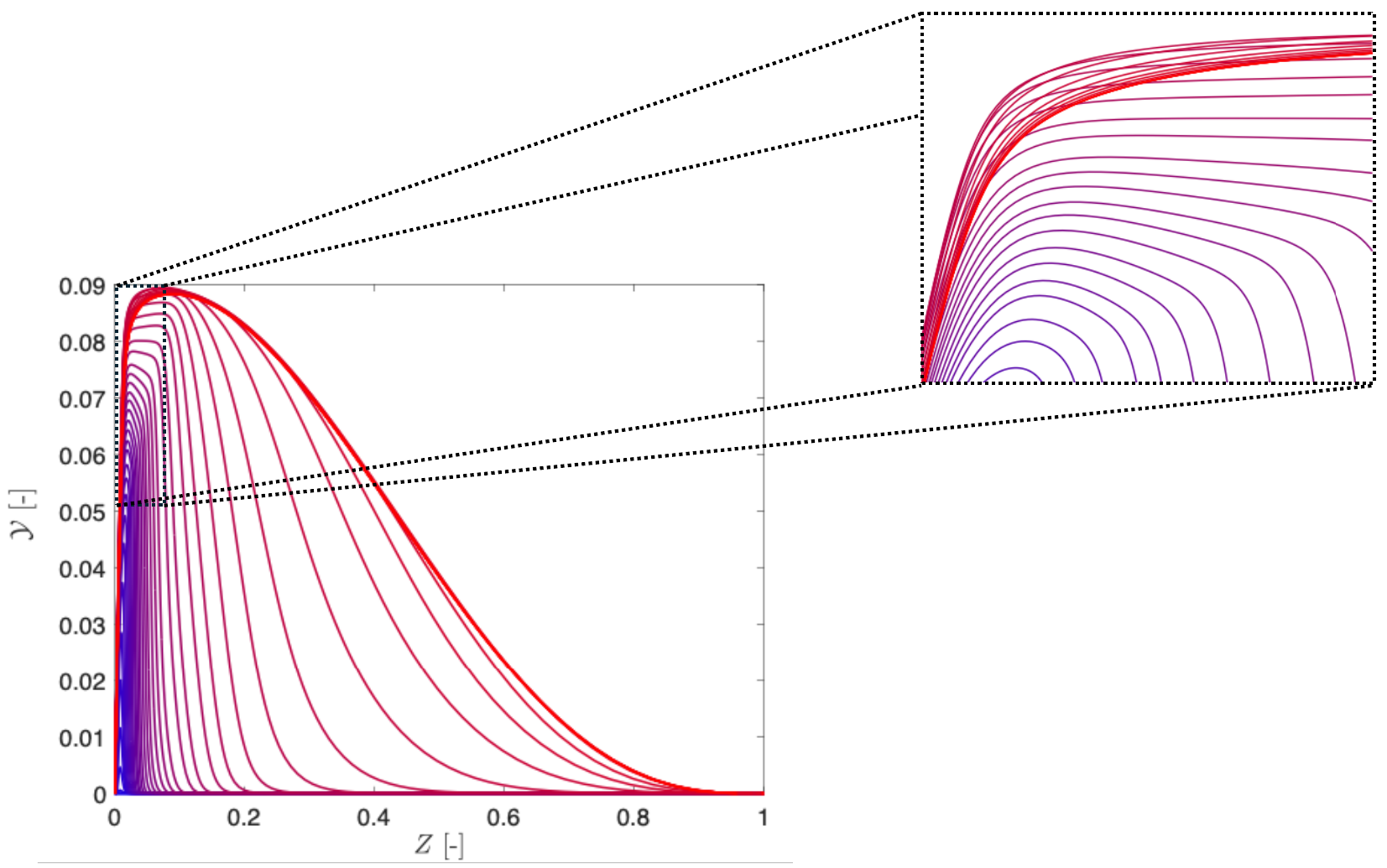}
    \caption{Progress variable $\mathcal{Y} = Y_{\ce{H2O}} - 10Y_{\ce{H}}$ as function of mixture fraction $Z$ in 1D flamelets, zoomed in for $ 0.005 < Z < 0.05$ and $ 0.05 < \mathcal{Y} < 0.09$. The colour gradation from blue to red denotes the time advancement. For very lean conditions, the progress variable, after reaching a maximum, starts decreasing.}
    \label{nonMonotonic}
\end{figure}

\addvspace{10pt}
First, we define the control variables.
 For mixture fraction, a modified expression from Bilger \cite{Bilger} is adopted:

\begin{equation}
    Z = \frac{\frac{1}{2}\frac{Z_{H}-Z_{H,2}}{M_{H}} - \frac{Z_{O}-Z_{O,2}}{M_{O}}}{\frac{1}{2}\frac{Z_{H,1}-Z_{H,2}}{M_{H}} - \frac{Z_{O,1}-Z_{O,2}}{M_{O}}}
\end{equation}
where $Z_{i}$ denotes the elemental mass fraction and $M_{i}$ the molar mass fraction of hydrogen and oxygen atoms. The subscripts 1 and 2 refer to, respectively, the fuel and the oxidizer stream. 
The progress variable $\mathcal{Y}$ is defined as follows:

\begin{equation}
\label{pv}
    \mathcal{Y} =  \sum_{i}\alpha_{i}Y_{i}
\end{equation}
where $\alpha_{i}$ and $Y_{i}$ denote the weight factor and the mass fraction of the i$^{th}$ species, respectively.
As mentioned, the high diffusivity of \ce{H2} and \ce{H} radicals often prevents the progress variable, as defined in Eq.~(\ref{pv}), to satisfy such requirement in non-premixed igniting flamelets. For the case investigated in the present study, this non-monotonicity on the very lean side hinders a correct description of the quasi-steady burning process (Fig.~\ref{nonMonotonic}). After the ignition has taken place, the progress variable starts diffusing from lean towards richer composition. This means that, for a certain range in $Z$, the progress variable $\mathcal{Y}$ is not monotonic, as it first increases during ignition and then decreases during the quasi-steady burning phase.

To avoid this problem, we decided to replace the common parametrisation based on diffusion flamelets with a tabulation stemming from 0D homogeneous reactors (HR-FGM). We start by computing an inert counterflow flamelet with strain rate $a = \SI{100}{\per\s}$, which is a common value used extensively in literature (e.g. \cite{wang}). The boundary conditions for species mass fractions and temperature ($T$) are reported in Table \ref{bcflam}, with a stoichiometric mixture fraction $Z_{st} = 0.0102$. Differential diffusion is here accounted for by assuming constant Lewis numbers $Le_{i}$. Using the compositions along this flamelet as initial conditions, a series of perfectly stirred reactors at constant-pressure is computed in the range $0 < Z < Z_{R}$ and used to generate the first part of the table. 

\begin{table}[b!] \footnotesize
\caption{Boundary conditions.}
\centerline{\begin{tabular}{lcc}
\hline 
                     & Fuel  & Oxidizer      \\
\hline

$T$ (K)                               & 300 & 1300          \\
$Y_{\ce{H2}}$ (-)    & 1  & 0      \\ 
$Y_{\ce{O2}}$ (-)    & 0  & 0.082      \\ 
$Y_{\ce{Ar}}$ (-)    & 0  & 0.918      \\ 
\hline 
\end{tabular}}
\label{bcflam}
\end{table}

The second part of the table, i.e. for $Z_{R}<Z<1$, needs special treatment because homogeneous reactors do not ignite. Three methods are explored:
 
\begin{itemize}
    \item The most straightforward approach is based on assuming that the species mass fractions $Y_{i}$ and the enthalpy $h$ follow an inert mixing rule with unity Lewis number diffusion. Hence, they are linear functions of $Z$, with temperature and density computed from such ($Y_{i}$, $h$) states. This method is based on inaccurate assumptions, so it is expected to behave poorly. Nevertheless, it is a widely used approach in unity-$Le$ type of flames, so it is worth an investigation. This method will be referred to as the HRIM-FGM. 
    \item The second method is to force ignition of the reactors by adding a spurious source of reaction to the 0D model, such that the mixture in the reactor is driven to high enough temperature to eventually ignite. In practice, this is achieved by adding the global reaction 2\ce{H2} + \ce{O2} $\to\,$ \ce{H2O} with low energy of activation such that it occurs at low temperature values, but also low rate such that the influence on the chemistry after ignition is small. The source term and heat release rate of this spurious reaction are excluded when tabulating the thermochemical properties. This method will be referred to as the HRFI-FGM and represents an improvement with respect to the HRIM-FGM, as it accounts for preferential diffusion along the entire mixing line, but it still based on unphysical assumptions. On the other hand, it can be an interesting approach since it is completely based on homogeneous reactors.
    \item The last approach leverages on the diffusion flamelets in the region where the progress variable is monotonic. In particular, it uses a time-series of igniting diffusion flamelets (HRDF-FGM) evolving from the starting solution of the mixing line, plus steady counterflow flamelets with decreasing strain rate \cite{vanoijen}. Because of the non-matching values at the interface $Z = Z_{R} = 0.045$ between the tabulation with 0D reactors and the one with 1D flamelets, a weight function is used to smoothen the transition between the two, namely $W = \frac{1}{2}[1+\mathrm{tanh}(\frac{Z-Z_{R}+3dZ}{dZ})]$, where $dZ = 0.005$ specifies the transition width between the reactors and the flamelets. This operation leads to an arbitrary modification of the tabulated information, that is, however, confined to a limited area in mixture fraction space.
\end{itemize}

\begin{figure}
\centering
         \begin{minipage}[h!]{1\textwidth}
        \includegraphics[width=1\linewidth]{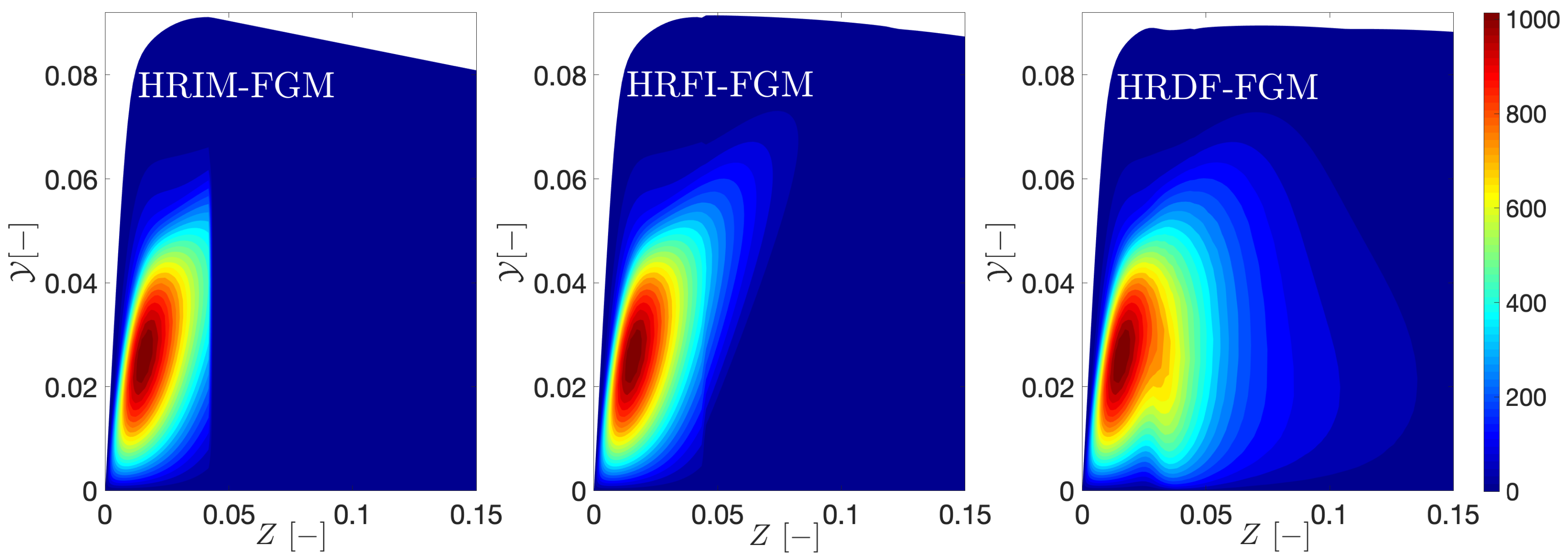} 
        \label{HRD}
     \end{minipage}
\centering
\hfill
\caption{The source of progress variable [kg/m$^{3}$s] in $Z$-$\mathcal{Y}$ space for the different extrapolation methods, zoomed in on  $Z \leq$ 0.15, with $Z_{R}$ = 0.045.}
\label{fgms}
\end{figure}

With respect to Eq.~(\ref{pv}), we found that a combination of \ce{H2O} and \ce{H} mass fractions with $\alpha_{\ce{H2O}} = 1$ and $\alpha_{\ce{H}} = -10$ yields a suitable progress variable for the HR-FGM. The resulting tables are shown in Fig.~\ref{fgms}. The source of progress variable in $Z$ $<$ $Z_{R}$ is identical for all the three methods since they rely on the same tabulation based on homogeneous reactors. The HRIM-FGM has an abrupt jump in the source of progress variable due to the assumption of inert mixing and therefore $\omega_{\mathcal{Y}}$ = 0. The linearity of the extrapolation method is visible in the maximum progress variable for $Z \in [Z_{R}, 1]$. The HRFI-FGM is smoother in the source of progress variable thanks to the forced ignition of the inert reactors. Still, the source of $\mathcal{Y}$ is non-zero for $\mathcal{Y} >$ 0.02 only. Moreover, the source of progress variable of these reactors is close to zero for $Z >$ 0.08, due to the missing diffusive effects in the mixture fraction direction. The maximum progress variable of this method corresponds to the chemical equilibrium. The HRDF-FGM takes advantage of the corresponding 1D flamelets in the region where the progress variable is monotonic. Thanks to the diffusion of the reactive species from the lean side, the source of progress variable covers a greater portion of the $Z$-$\mathcal{Y}$ space. 

\section{1D counterflow simulations}\addvspace{10pt}

\subsection{Numerical}\addvspace{10pt}

First, we want to assess the validity of the different FGM tabulation methods in one-dimensional flame configurations. To conduct such comparison, we use the CHEM1D software. For details on the implementation and the equations that are solved, one can refer to \cite{somers}.

We assume a counterflow type of flame, with a cold fuel stream of hydrogen and a hot oxidizer stream of argon-oxygen at constant pressure $p = 1$ bar and strain rate $a = \SI{100}{\per\s}$. The boundary conditions for the species and temperature are the same as in Table~\ref{bcflam}. The results of the different FGM methods are compared against computations with detailed chemistry (9-species, 19-reactions mechanism from Burke et al.~\cite{burke}).

\subsection{Results}\addvspace{10pt} \label{results1d}

\begin{figure}[t!]
\centering
\includegraphics[width=0.8\linewidth]{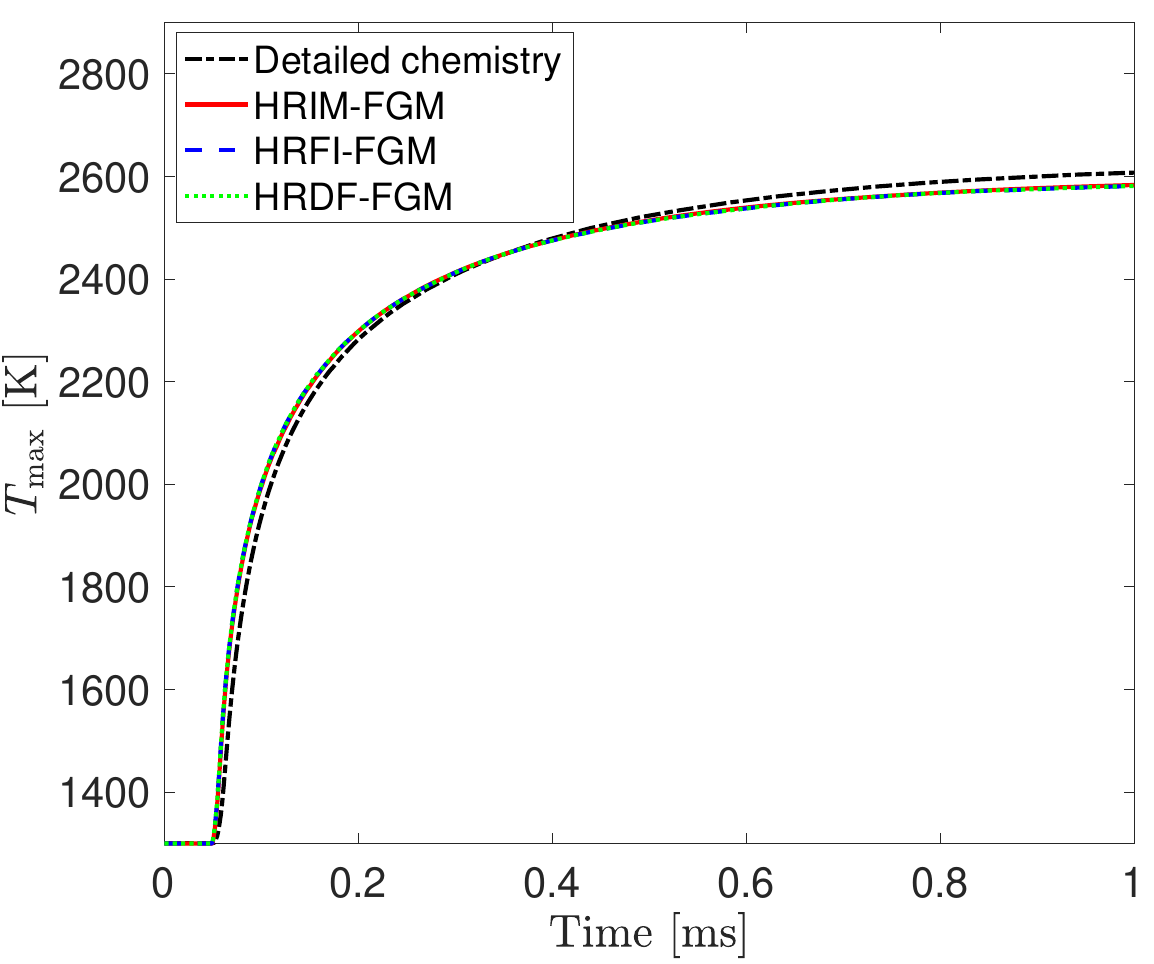} 
\caption{Temperature as function of time for flamelets computed with detailed chemistry and with different HR-based FGM tables.}
\label{igndel}
\end{figure}

\begin{figure}[t!]
\centering
\includegraphics[width=1.0\linewidth,trim={7cm 2.5cm 7cm 1.5cm},clip]{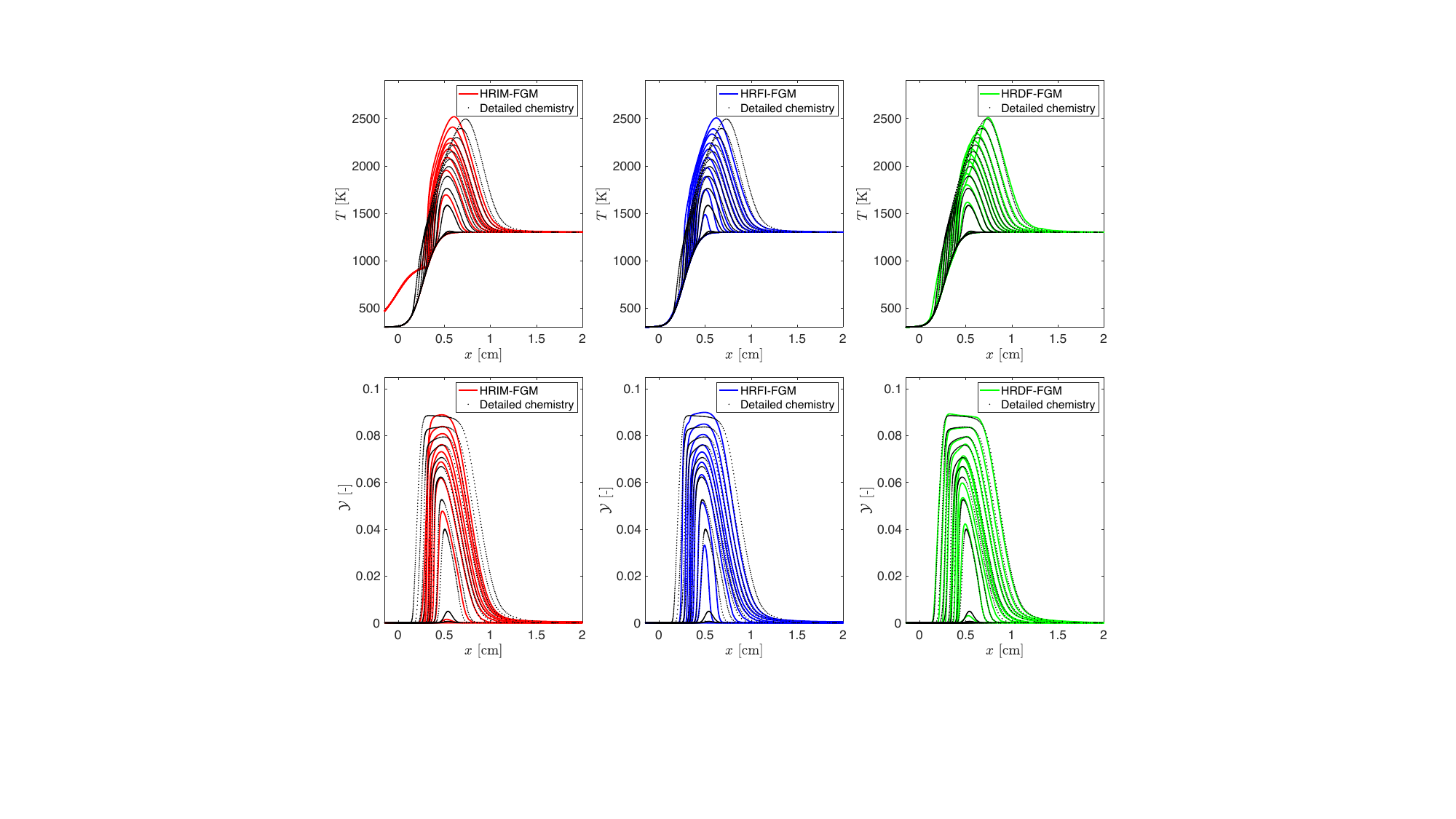} 
\caption{Time evolution of temperature (top) and progress variable (bottom) as function of the $x$ coordinate for the three different HR-based FGM methods.}
\label{pvxflam}
\end{figure}

Figure \ref{igndel} reports the time evolution of the maximum temperature $T_\mathrm{max}$ in the domain. Both the ignition delay (measured as the time at which $T_\mathrm{max}$ has increased by 100 K) and the evolution of the maximum temperature afterwards are captured well by all the three tabulation techniques. Moreover, the three FGM curves coincide, revealing that the ignition process is mainly determined by the homogeneous reactors in $Z$ $<$ $Z_{R}$, whose related tabulation is the same for all the three methods. Figure \ref{pvxflam} shows the spatial behaviour of temperature and progress variable in time, respectively. The HRIM-FGM overpredicts the temperature at the fuel (left) side when compared to detailed chemistry, as it assumes unity Lewis numbers for a large part of the composition space. The HRFI-FGM, on the contrary, does not show such deviation as the mixing line beyond $Z_{R}$ is not extrapolated and accounts for differential diffusion impacting the extrapolated composition beyond the reactors’ flammability limit. However, both the HRIM-FGM and HRFI-FGM slightly underpredict the flame thickness as their contribution to the source of progress variable in the $Z$ $>$ $Z_{R}$ part of table is marginal, whereas the HRDF-FGM follows the detailed chemistry results accurately.

\begin{figure*}
    \centering
    \includegraphics[width=1\linewidth]{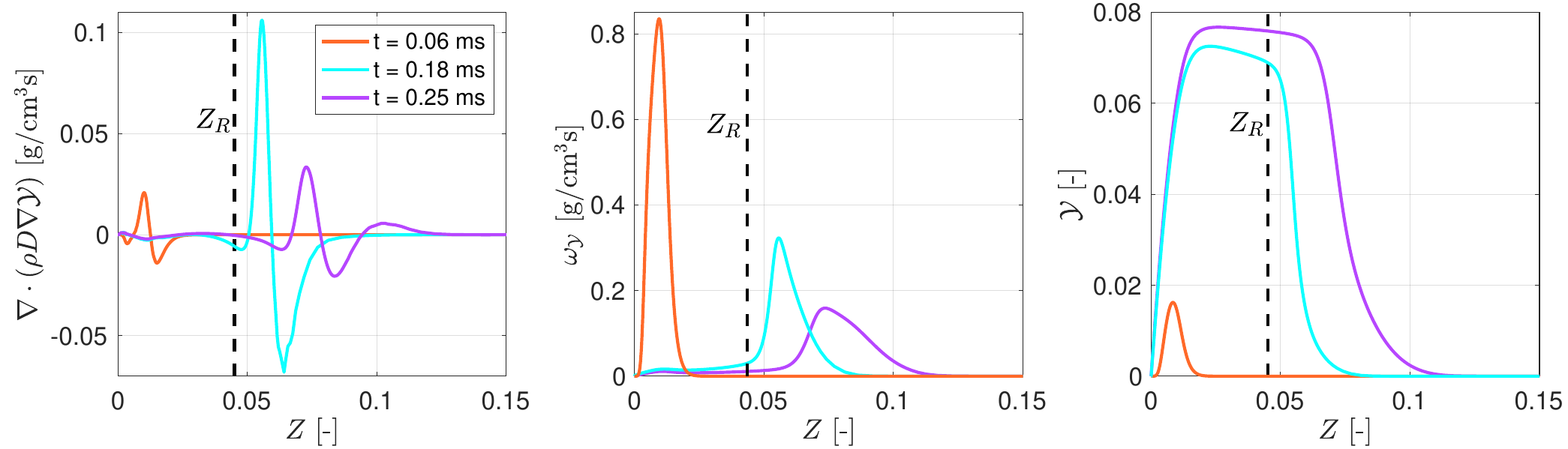}
    \caption{Budget analysis in 1D flame configuration as function of mixture fraction with detailed chemistry.}
    \label{fluxes}
\end{figure*}

To further demonstrate the validity of the HRDF-FGM, we performed a budget analysis of the diffusive and chemical terms in the balance equation for $\mathcal{Y}$ (Fig.~\ref{fluxes}). As abovementioned, the tabulated chemistry based on homogeneous reactors has already been employed in literature, but mainly for spray flames (e.g. \cite{lucchini}), where $Le$ = 1 can be assumed for all the species. In that context, the extrapolation method of the manifold has little impact on the overall combustion process, since $Z$ is hardly higher than the flammability limit.
This is not the case for the work at hand. The progress variable $\mathcal{Y}$ together with the corresponding diffusive flux term $\nabla\cdot(\rho D \nabla \mathcal{Y})$ and chemical source term $\omega_{\mathcal{Y}}$ computed with detailed chemistry are reported at three time instances as function of mixture fraction in the 1D flame configuration. When ignition occurs ($t$ = 0.06 ms), it is visible that chemistry dominates the evolution of $\mathcal{Y}$ for $Z$ $<$ $Z_{R}$, as $\omega_{\mathcal{Y}}$ is almost two orders of magnitude larger than the corresponding diffusive flux term. This justifies the use of homogeneous reactors at the lean side, since the loss of diffusive effects in the tabulation can be deemed negligible. On the other hand, for $Z_{R} < Z < 0.15$, $\omega_{\mathcal{Y}}$ and $\nabla\cdot(\rho D \nabla \mathcal{Y})$ are comparable in magnitude, both for $t$ = 0.18 ms and $t$ = 0.25 ms, and cannot be neglected. This is why the HRIM-FGM and the HRFI-FGM start to be inaccurate: the former does not include any chemical source term, the latter includes chemistry but no diffusivity of the species in mixture fraction space. In particular, the differences shown in Fig. \ref{pvxflam} between the HRFI-FGM and the HRDF-FGM are mainly related to the flame position, that looks shifted in the HRFI-FGM. However, this regards the high-temperature lean region, where the tabulation of the HRFI-FGM and the HRDF-FGM is identical, since it is based on the same homogeneous reactors in both cases. The reason for this difference has then to be found in the extrapolated side of the table. Indeed, for $x <$ 0.3 cm, the flame as predicted by the HRFI-FGM does not react as expected. This is due to the fact that, for $Z >$ 0.07, the HRFI-FGM has no source of progress variable, as it does not account for diffusive effects in the reaction. This unchanged composition at the richer side does not allow the flame to diffuse correctly, contrarily to the HRDF-FGM where the reaction layer well agrees with the corresponding detailed chemistry. Consequently, an extrapolation based on 1D flamelets is deemed needed to account both for chemistry and diffusivity in the slightly richer composition.

\begin{figure}[h!]
    \centering
    \includegraphics[width=0.8\linewidth]{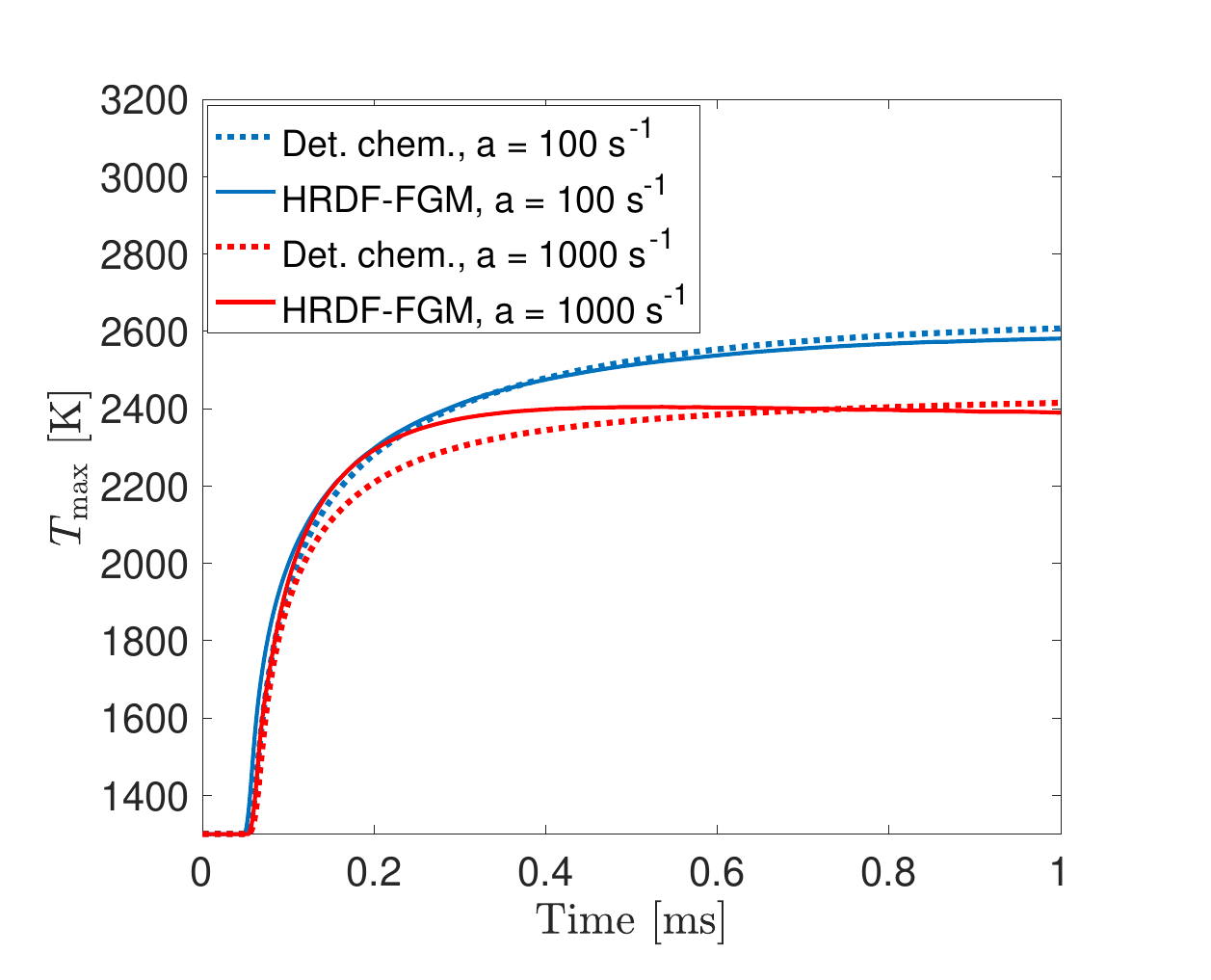}
    \caption{Maximum temperature [K] as function of time for different strain levels as predicted by the HRDF-FGM.}
    \label{strain}
\end{figure}

In order to assess the sensitivity to scalar dissipation, the same HRDF-FGM manifold based on flamelets with a strain rate of $a$ = 100 s$^{-1}$ is applied to a counterflow flame with a ten times higher strain level of $a = \SI{1000}{\per\s}$ and the results are compared with the corresponding detailed chemistry results (see Fig.~\ref{strain}). It is visible how the proposed method is capable of accurately describing both the ignition delay and the following quasi steady-state phase in a proper way, although differences can be observed around \SI{0.2}{\ms}. It is worth to notice that the ignition delay looks insensitive to strain rate, confirming that the diffusion processes may not play a significant role in this phase.

\section{LES of turbulent mixing layer}\addvspace{10pt}

Because of its best performance, the proposed HRDF-FGM approach is now assessed via LES of 3D turbulent reacting mixing layers, and the results are validated against DNS with detailed chemistry. For consistency, the mechanism of Burke et al. has been employed in the DNS as well.

\subsection{Numerical approach}\addvspace{10pt}

The governing equations for LES are the filtered Navier-Stokes for mass and momentum conservation, plus the filtered transport equations for the control variable $\phi = (Z, \mathcal{Y})$, namely:

\begin{equation}
    \frac{\partial(\overline{\rho}\widetilde{\phi})}{\partial t} + \nabla\cdot(\overline{\rho}\widetilde{\bm{u}}\widetilde{\phi}) - \nabla\cdot\Bigg[\Bigg(\frac{\lambda}{c_{p}} +\frac{\mu_{t}}{Sc_{t}}\Bigg)\nabla\widetilde{\phi}\Bigg] = \nabla\cdot(d_{\phi}\nabla\widetilde{Z}) + \overline{{\omega_{\phi}}}
\end{equation}
where the symbols $\bar{.}$ and $\widetilde{.}$ denote the Reynolds and Favre averaged quantities, respectively; $\rho$ is the gas density and $u$ is the flow velocity. The last term on the left-hand side is the sum of the laminar and the turbulent diffusivity, respectively. In particular, $\lambda$ is the thermal conductivity, $c_{p}$ is the heat capacity at constant pressure, $\mu_{t}$ is the turbulent viscosity and $Sc_{t}$ is the turbulent Schmidt number, here assumed to be equal to 0.7 \cite{spalding}. The first term on the right hand side denotes the constant-Lewis preferential diffusion term, with $d_{\phi}$ being the corresponding diffusion coefficient \cite{vanoijen2}, whereas $\overline{{\omega_{\phi}}}$ is the source term, equal to 0 for $\phi = Z$.

To account for turbulence-chemistry interaction, a presumed $\beta$-PDF (Probability Density Function) is assumed for mixture fraction \cite{bouras}. A third controlling variable, the variance of mixture fraction $\widetilde{Z''^{2}}$, is then taken into account by solving:

\begin{equation}
\label{eqVar}
    \frac{\partial(\overline{\rho}\widetilde{Z''^{2}})}{\partial t} + \nabla\cdot(\overline{\rho}\widetilde{\bm{u}}\widetilde{Z''^{2}}) -\nabla\cdot\Bigg[\Bigg(\frac{\lambda}{c_{p}} + \frac{\mu_{t}}{Sc_{t}}\Bigg)\bm\nabla\widetilde{Z''^{2}}\Bigg]  =  
     \\ 2\Bigg[\Bigg(\frac{\lambda}{c_{p}} + d_{Z} + \frac{\mu_{t}}{Sc_{t}}\Bigg)[\bm{\nabla}\widetilde{Z}]^2-2\overline{\rho}C_{d}\frac{\mu_{t}}{Sc_{t}}\frac{\widetilde{Z''^{2}}}{\Delta^{2}}\Bigg] 
\end{equation}
where the constant $C_{d}$ is equal to $\frac{1}{4}$ and $\Delta$ is the filter width (here the cell volume $V_{c}^{1/3}$). The simulations have been carried out within the OpenFOAM framework \cite{bao}, employing an implicit Euler backward scheme for time integration and a cubic convective scheme for the divergence terms. The sub-grid scale (SGS) closure is characterized by a dynamic $k$-equation model \cite{ballatore}. 

The LES results are validated against DNS with detailed chemistry. In particular, an in-house, fully compressible, compact finite difference scheme with fifth-order accuracy in space, and third-order accuracy in time has been employed. The implementation of the boundaries employs Navier-Stokes characteristic boundary conditions (NSCBC) with sub-sonic, partially reflecting outlet, with convective and acoustic transverse terms treatment in the $y$-$d$ direction \cite{Couss}. For more details, please refer to \cite{diego}.

\subsection{Case set-up} \addvspace{10pt}

\begin{figure}[t!]
    \centering
    \includegraphics[width = 0.8\linewidth]{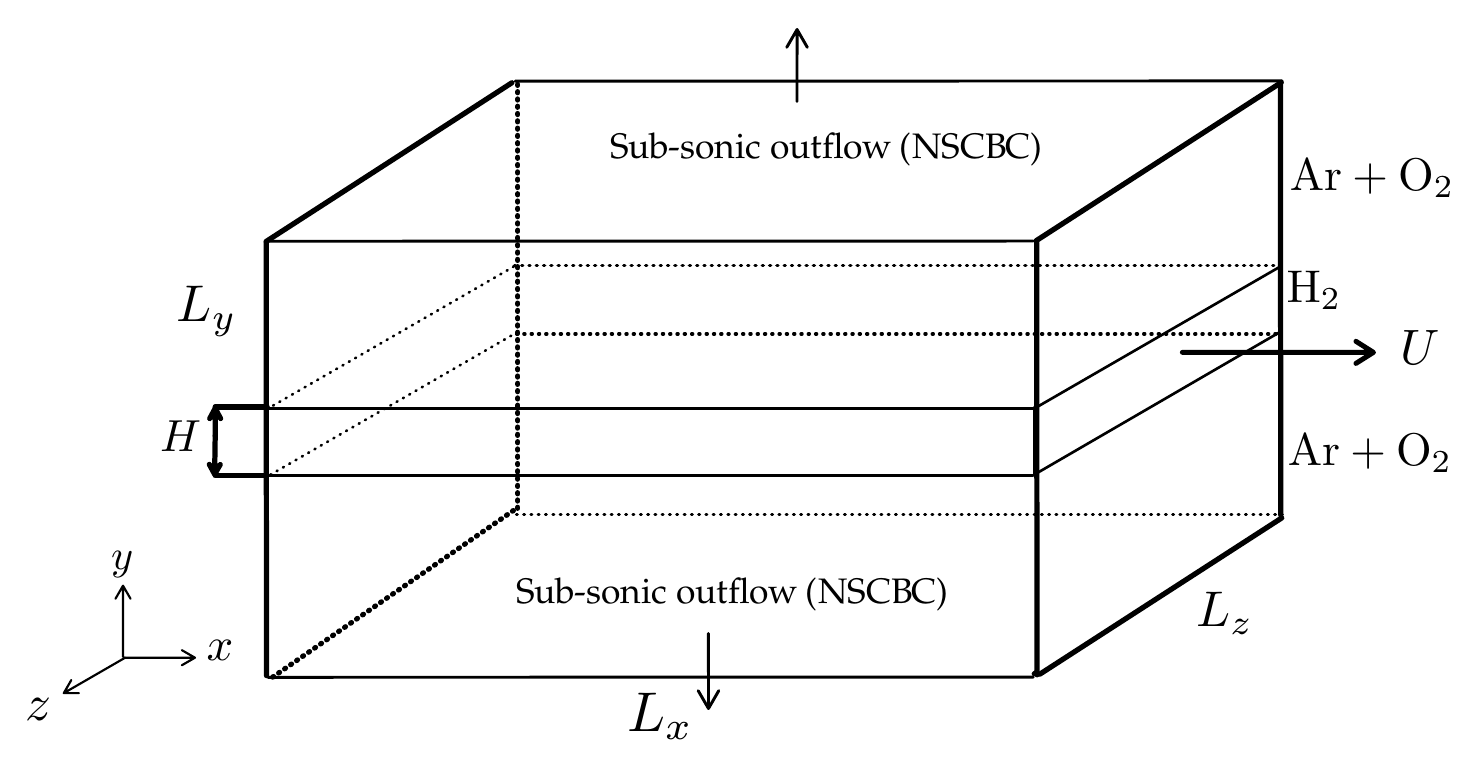}
    \caption{Schematic representation of the domain of the turbulent mixing layer.}
    \label{schemMix}
\end{figure}

The turbulent mixing layer set-up is schematically depicted in Fig.~\ref{schemMix}. The domain is periodic in $x$ and $z$ direction, and has outflow conditions in $y$-direction. The initial solution for LES is filtered from the DNS and interpolated onto the LES mesh. In particular, the initial velocity is defined with a hyperbolic tangent function that approximates a top-hat profile centered at the $y$ plane, that ranges from 0 to the maximum value $U$ in a thickness of $d$ = 2 mm. The jet Reynolds number is $Re$ $= HU/\nu_{\ce{H2}}$ = $10^{4}$, with $U$ = 565 m/s, $H$ = 2 mm, and $\nu_{\ce{H2}}$ being the kinematic viscosity of hydrogen. The initial turbulence $u_i^\prime$ is created using a von-Karman-Pao spectrum with a root mean square value for the fluctuations equal to 56.5 m/s (i.e. 10\% turbulence intensity). The jet core and surrounding environment have the same composition and temperature of the fuel and oxidizer, respectively, as reported in Table \ref{bcflam}.
The domain has been discretised both in DNS and LES with a uniform mesh along the three spatial directions ($L_{x} = 15$ mm, $L_{y} = 12.5$ mm, $L_{z} = 7.5$ mm), with different mesh sizes (see Table \ref{meshes}).

\begin{table} \footnotesize
\caption{Mesh width $\Delta$ for DNS and the three LES cases.}
\centerline{\begin{tabular}{lc}
\hline 
                     & $\Delta$ (mm)     \\
\hline
DNS                                  & 0.0125           \\
LES, fine                               & 0.0500          \\
LES, medium                               & 0.0875          \\
LES, coarse                               & 0.1250          \\
\hline 
\end{tabular}}
\label{meshes}
\end{table}

\subsection{Results} \addvspace{10pt}

\begin{figure}[t!]
\centering
\includegraphics[width=1\linewidth]{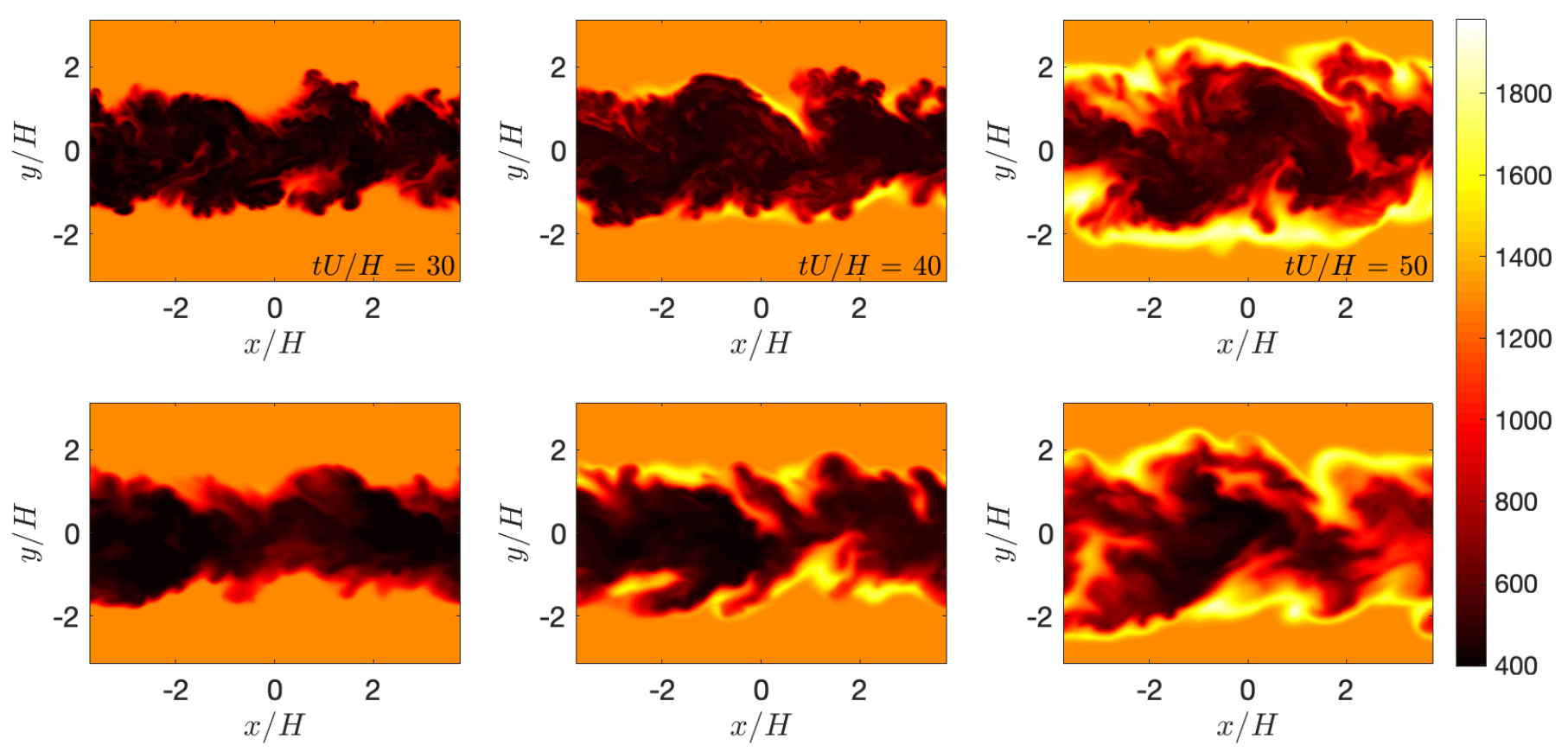} 
\caption{Mid-plane instantaneous contours of temperature [K] at jet times $tU/H = 30, 40, 50$, from left to right respectively. Top: filtered DNS, bottom: LES, fine.}
\label{contours}
\end{figure}

A first overview of the evolution of the hydrogen flame in the mixing layer as predicted by DNS and HR-FGM-LES with the fine mesh is reported in Fig.~\ref{contours} at three different time instances, namely $tU/H$ = 30, 40, 50, respectively. The time is here normalised with the initial mean velocity and core width, i.e. the jet time $t_{jet} = H/U = 3.5$ $\mu$s. The DNS solution has been filtered with the same LES filter width to have a fairer comparison.  In general, the global trend is very similar: both the ignition and the burning processes are globally well captured by the proposed HR-FGM-LES approach. The hydrogen starts mixing with the surrounding environment and spreads along $y$ ($tU/H$ = 30). When ignition occurs, the flame starts growing along the stoichiometric mixture fraction isocontour ($tU/H$ = 40) and the hydrogen core broadens more rapidly in jet-normal direction ($tU/H$ = 50).

\begin{figure}
\centering 
\includegraphics[width=0.8\linewidth]{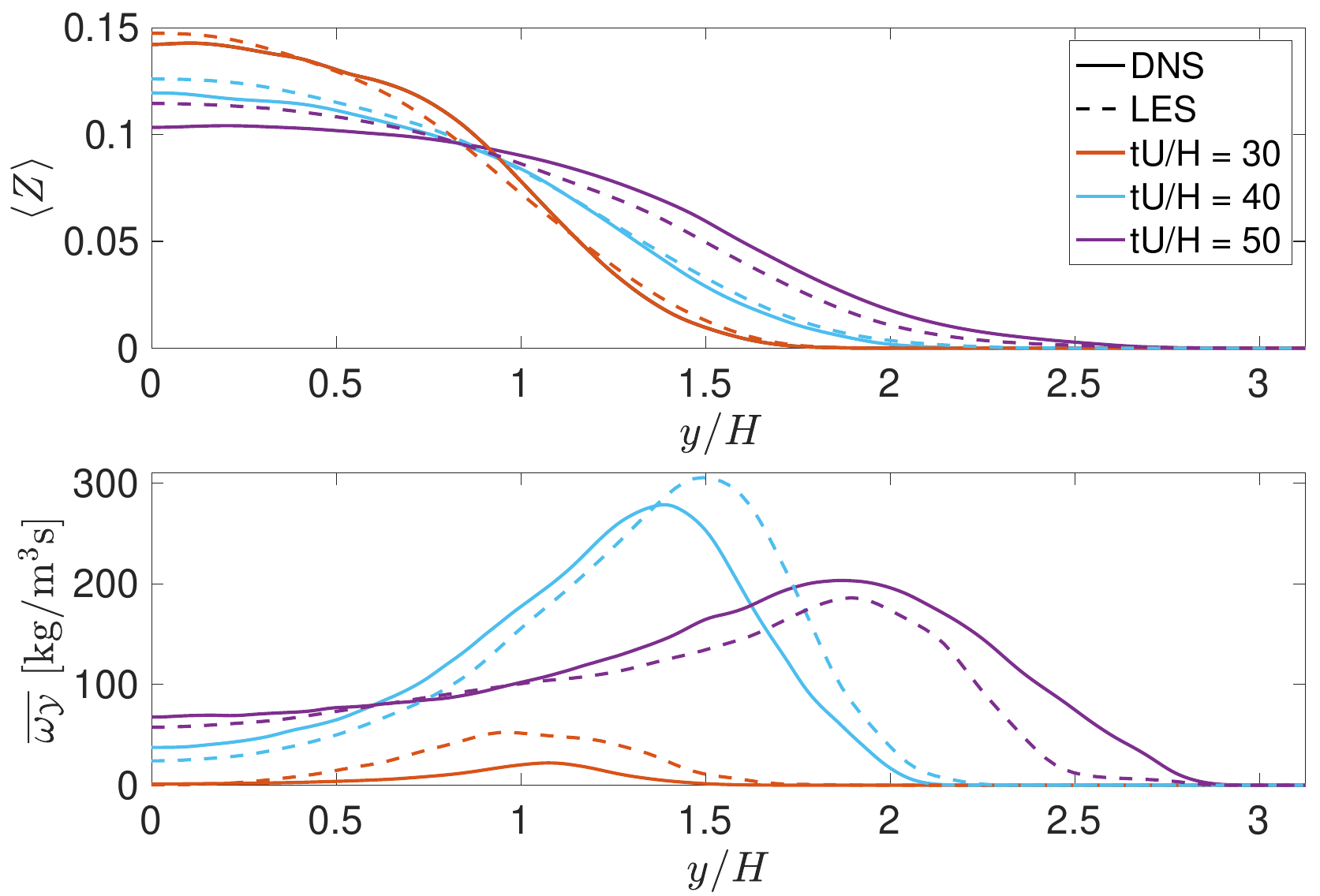} 
\includegraphics[width=1\linewidth]{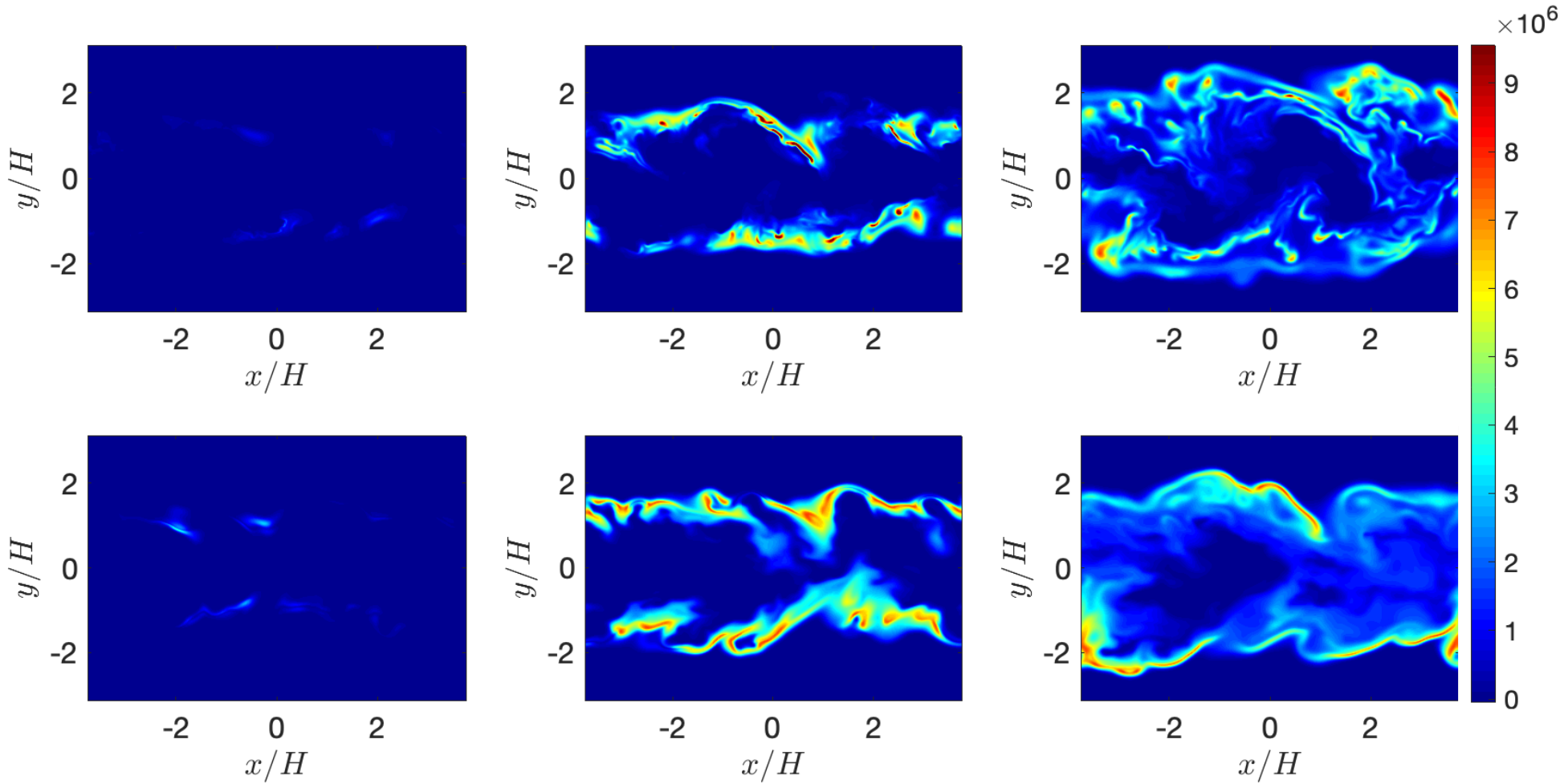}
\caption{Left: Favre-averaged mixture fraction and Reynolds-averaged source of progress variable profiles as function of $y/H$ at three different jet time instances $tU/H = 30, 40, 50$. Right: heat release rate [kW/$\mathrm{m^{3}}$] contours for the same times; top: filtered DNS, bottom: LES.}
\label{mixingandsrc}
\end{figure}

Figure \ref{mixingandsrc} reports the Favre-averaged mixture fraction $\langle Z \rangle$ and the Reynolds-averaged source of progress variable $\overline{\omega_{\mathcal{Y}}}$, averaged along the statistically-independent directions $x$ and $z$, as function of the $y$ coordinate at three different time instances. Also, the contours of heat release rate are reported for the DNS and LES for the same jet times. It is visible how the $\langle Z \rangle$ profiles are quite close to each other. Still, there are some differences. In particular, the jet as predicted by the LES spreads slightly more slowly than the DNS between jet times 40 and 50. This reflects in the source of progress variable. In fact, at the last reported time instance, $\overline{\omega_{\mathcal{Y}}}$ as predicted by the LES is lower than in the DNS, especially for $y/H >$ 1. This means that, due to the different mixing, the reaction is on average slightly less intense and less spread in space. The same considerations can be drawn when looking at the heat release rate contours. Indeed, since ignition is slightly anticipated, the jet releases more heat in the LES than in the DNS at $tU/H$ = 30 and 40, whereas at $tU/H$ = 50 a greater portion of the domain does not react because the jet is less spread in $y$ direction. Further insights on the temporal evolution of the heat release rate are provided in the following.

\begin{figure}[t!]
    \centering
    \includegraphics[width=1\linewidth]{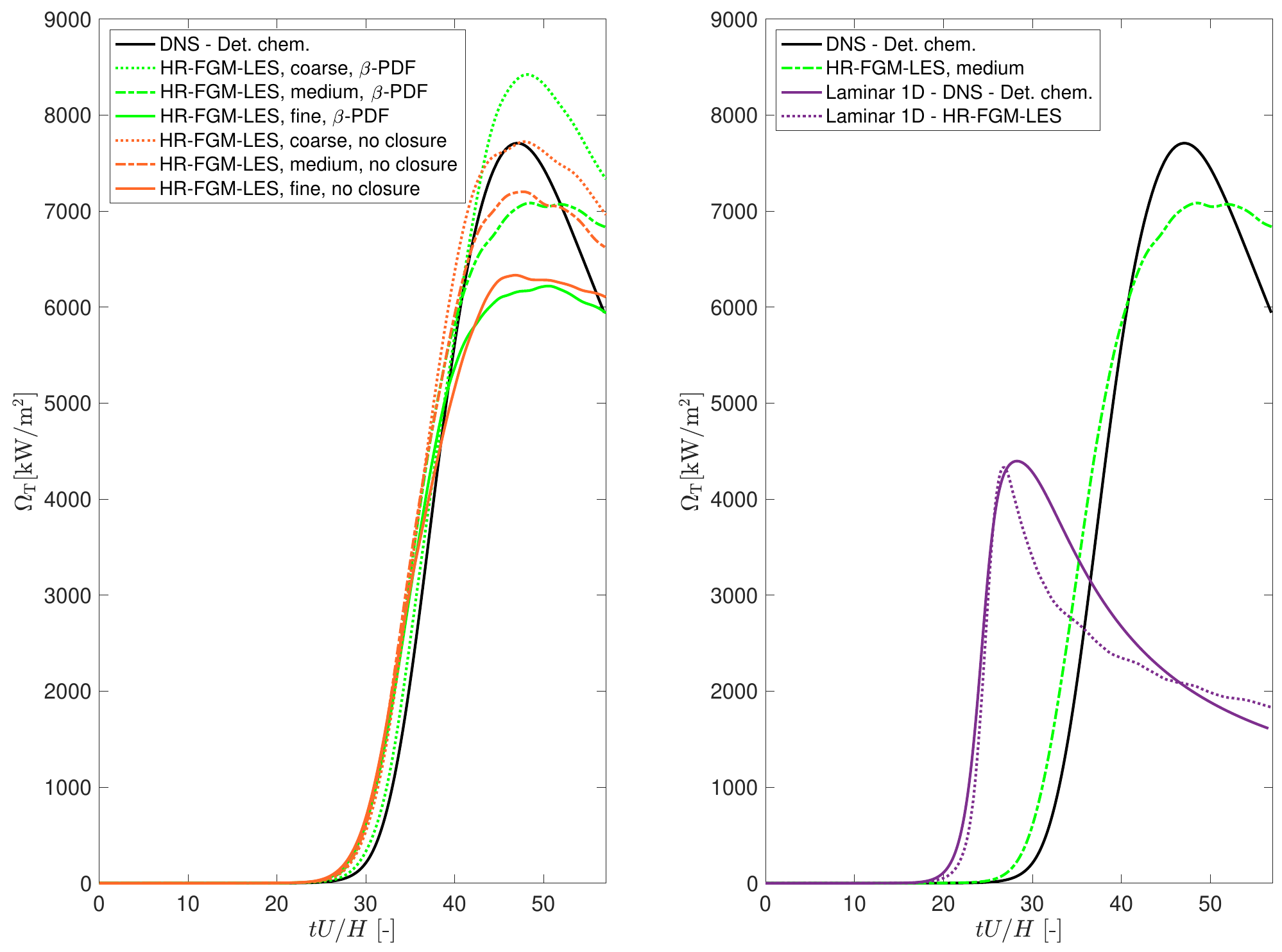}
    \caption{Integrated heat release rate $\Omega_{T}$ as function of time. On the left: different filter widths, with and without turbulence-chemistry interaction. On the right: comparison with 1D mixing layer.}
    \label{newhrr}
\end{figure}

Figure \ref{newhrr} shows the evolution of the spatially integrated heat release rate per unit of averaged flame surface $\Omega_{T}$ as function of time:

\begin{equation}
    \Omega_{\mathrm{T}} = \frac{1}{2L_{x}L_{z}}\int_{V}\omega_{\mathrm{T}}dV
\end{equation}
where $\omega_{\mathrm{T}} = -\sum_{i}{h_{i}\omega_{i}}$ (LES) and $\omega_{\mathrm{T}} = -\sum_{i}{(h_{i}-RT)}\omega_{i}$ (DNS) is the heat release rate, with $h_{i}$ and $\omega_{i}$ being the enthalpy and the chemical source term of the $i^{th}$ species, respectively, and $V$ is the volume of the domain. Since the DNS was run for 57 jet times only, the simulation time of the LES has been limited accordingly. 
The left plot of Fig.~\ref{newhrr} shows the integrated heat release rate for different mesh sizes, with and without turbulence-chemistry interaction. The global trend resembles the one of the DNS. The flame ignites slightly earlier than in the DNS, but the difference can be deemed small (around 6\%, corresponding to 7 $\mathrm{\mu}$s). The ignition delay (here measured as the time instance at which $\Omega_{\mathrm{T}}$ has increased to 200 kW/$\mathrm{m^{2}}$) is little to none affected neither by the filter width nor the $\beta$-PDF modelling, meaning that it is mainly driven by the source of progress variable as stored in the table. Moreover, it is visible that the impact of the PDF is reduced as the mesh gets finer. This is in line with theory, as the turbulent flow structures are more resolved when the filter width is smaller. In order to understand whether the differences with respect to the DNS are effectively due to the filter width, or rather to a different description of turbulence, we look at the $\Omega_{\mathrm{T}}$ of a laminar 1D mixing layer, with initial zero velocity, for both DNS and LES medium mesh (see second plot of Fig.~\ref{newhrr}). In both cases, the ignition delay is sensibly shortened when $U$ = 0 m/s, i.e. when there are no multi-dimensional flame effects. Also, the heat release rate increases much more rapidly, and the peak is remarkably decreased compared to the 3D case. This might be the effect of an increased flame curvature. Moreover, in presence of differential diffusivity, the flame dynamics is altered by flame curvature \cite{goktolga}. Such trend is overall captured quite nicely by the proposed HR-FGM method. Furthermore, the differences between detailed chemistry and HR-FGM in ignition delay and $\Omega_{\mathrm{T}}$ peak are sensibly reduced in 1D, even though the HR-FGM-LES mesh is 7x larger. This shows that the filtering operation has little impact on the flame dynamics. 

\begin{figure}[t!]
    \centering
    \includegraphics[width=1\linewidth]{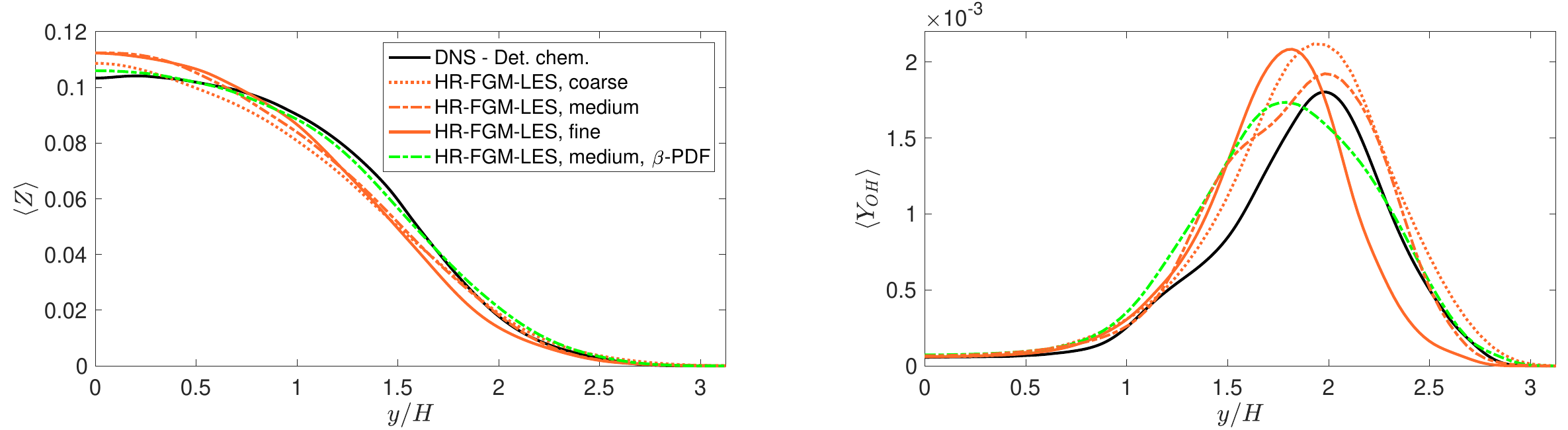}
    \caption{Favre-averaged  mixture fraction (left) and \ce{OH} mass fraction (right) profiles for different grid sizes and turbulence-chemistry interaction.}
    \label{oh}
\end{figure}

Finally, we study the flame properties in space, focusing on the impact of the mesh size and turbulence-chemistry interaction on the LES solution. Since the \ce{OH} mass fraction is known to be a flame marker, we look at Favre-averaged $Y_{\ce{OH}}$ profiles to qualitatively determine the reaction layer thickness and location. Figure \ref{oh} reports the $\langle {Y}_{\ce{OH}} \rangle$, as well as the corresponding $\langle Z \rangle$ profiles, as function of $y/H$, at $tU/H = 50$ for different grid sizes and turbulence-chemistry assumption.
Overall, the flame thickness and location is fairly consistent with respect to the corresponding DNS. The effect of a decreasing filter width is seen in a slower spreading of the mixture fraction, which reflects on the position of the reaction layer. In fact, the flame looks slightly narrower and more displaced towards the center of the jet. This might indicate that a turbulent Schmidt number lower than 0.7 is needed. The impact of the turbulence-chemistry interaction by means of the $\beta$-PDF is visible in a less irregular \ce{OH} and mixing profile. In particular, the $\langle {Y}_{\ce{OH}} \rangle$ looks smoother for richer compositions, i.e. for $|y|/H < 2$.

\section{Conclusions}
\addvspace{10pt}

The present study aimed at investigating turbulent auto-igniting hydrogen jets by means of tabulated chemistry, including preferential diffusion. Since the high diffusivity of hydrogen severely complicates the definition of a monotonic progress variable for the Flamelet-Generated-Manifold method, different possibilities to build a manifold based on 0D homogeneous reactors were assessed to overcome such problem. 

Different extrapolation methods were evaluated first in 1D flame configurations for validation against detailed chemistry solutions. A combination of homogeneous reactors and 1D flamelets gave the best results. The related budget analysis further confirmed that the extrapolation with diffusion flamelets past the reactors' flammability limit is necessary to retain both the chemical and diffusive effects in the slightly richer compositions. Such method was used in a large-eddy simulation to investigate a turbulent auto-igniting hydrogen flame in an oxygen-argon environment. Good agreement was found in terms of ignition delay, steady-state burning process and flame structure. Further simulations were carried out to investigate the sensitivity of the LES solution to various factors such as filter width, turbulence-chemistry interaction, and flow velocity. 

All these analysis represent valuable information in view of incorporating auto-igniting non-premixed combustion modelling in more complex geometries and applications that use injection of hydrogen in a high temperature environment.

\section*{Declaration of competing interest} \addvspace{10pt}

The authors declare that they have no known competing financial interests or personal relationships that could have appeared to influence the work reported in this paper.

\section*{Acknowledgments} \addvspace{10pt}

This work is part of the project Argon Power Cycle with project number 17868 of the Vici research programme, which is (partly) financed by the Dutch Research Council (NWO).

\bibliographystyle{elsarticle-num-names}
\bibliography{PCI_LaTeX}

\end{document}